\documentclass[epjST]{svjour}
\usepackage{graphics}
\usepackage{graphicx}
\usepackage{xcolor}
\usepackage{epsf}
\usepackage{amsmath}
\usepackage{amssymb}
\usepackage[english]{babel}
\usepackage{bm}
\usepackage{times}
\usepackage[T1]{fontenc}
\usepackage{graphics}
\usepackage{color}
\usepackage{dcolumn}   
\usepackage{sidecap}
\usepackage{ulem}

\newcommand {\ff} {\ensuremath{{\mathbf f}}}
\newcommand {\cc} {\ensuremath{{\mathbf c}}}
\newcommand {\vv} {\ensuremath{{\mathbf v}}}
\newcommand {\PP} {\ensuremath{{\mathbf P}}}
\newcommand {\EE} {\ensuremath{{\mathbf E}}}

\newcommand {\rr} {\ensuremath{{\mathbf r}}}
\newcommand {\xxi} {\ensuremath{{\mathbf \xi}}}
\newcommand {\Xxi} {\ensuremath{{\mathbf \Xi}}}
\newcommand {\cci} {\ensuremath{{\mathbf \chi}}}
\newcommand {\pphi} {\ensuremath{{\mathbf \phi}}}
\newcommand {\Pphi} {\ensuremath{{\mathbf \Phi}}}

\newcommand {\zz} {\ensuremath{{\bf z}}}
\newcommand {\Hh} {\ensuremath{{\bf H}}}
\newcommand {\TT} {\ensuremath{{\bf T}}}
\newcommand{\eps}{\epsilon}

\newcommand{\bes}{ \begin{equation} \begin{split} }
\newcommand{\ees}{ \end{split} \end{equation} }

\newcommand{\ignore}[1]{}

\newcommand{\gray}[1]{{\color{gray}{#1}}}

\renewcommand{\emph}[1]{{\it #1}}
\begin{document}
\title{A Tutorial on Time-Evolving Dynamical Bayesian Inference}
\author{Tomislav Stankovski\inst{1} \and Andrea Duggento\inst{2} \and Peter V. E. McClintock\inst{1} \and Aneta Stefanovska\inst{1}\fnmsep\thanks{\email{aneta@lancaster.ac.uk}} }
\institute{Department of Physics, Lancaster University, Lancaster, LA1 4YB, United Kingdom \and
Medical Physics Section, Faculty of Medicine, Tor Vergata University, Rome, Italy}
\abstract{
In view of the current availability and variety of measured data, there is an increasing demand for  powerful signal processing tools that can cope successfully with the associated problems that often arise when data are being analysed. In practice many of the data-generating systems are not only time-variable, but also influenced by neighbouring systems and subject to random fluctuations (noise) from their environments. To encompass problems of this kind, we present a tutorial about the dynamical Bayesian inference of time-evolving coupled systems in the presence of noise. It includes the necessary theoretical description and the algorithms for its implementation. For general programming purposes, a pseudocode description is also given. Examples based on coupled phase and limit-cycle oscillators illustrate the salient features of phase dynamics inference. State domain inference is illustrated with an example of coupled chaotic oscillators. The applicability of the latter example to secure communications based on the modulation of coupling functions is outlined. MatLab codes for implementation of the method, as well as for the explicit examples, accompany the tutorial.
} 
\maketitle
\section{Introduction}
\label{s:introduction}

In simple terms science might be defined as the systematic observation and analysis of nature and of how natural processes evolve in time and space. Such processes can include the beating of the mammalian heart, the movements of the planets, or simply how human society works.  In all of these cases, physics tries to generate a model based on data collected over time -- a time interval that depends on how fast the processes occur, which may in some cases be over centuries and in others over seconds or  microseconds. Science attempts to develop the models that can link most comprehensively the causes and consequences of the processes in question. One of the most frequently-used approaches, and arguably the most useful one, is the Bayesian approach. It is based on Bayes' theorem, which is central to the inverse problem approach and to dynamical inference, seeking to answer the question:  given a series of data resulting from observations, what can we deduce about the nature of the system or the process that generated that data?

There are many different ways in which Bayes' theorem may be used, and different groups of methods exist for performing so-called ``Bayesian inference'' \cite{Toussaint:11,Arulampalam:02,Friston:02,Smelyanskiy:05a,Smelyanskiy:05b,Penny:09}. Of particular interest among these are the Bayesian methods for dynamical inference. They enable a dynamical model in terms of ordinary or stochastic differential equations to be inferred from the observed data. Dynamical Bayesian methods provide the basis for important signal processing techniques that have been applied to e.g.\ physics, biology, communications, and climate \cite{Smelyanskiy:05a,Smelyanskiy:05b,Friston:02,Stankovski:14a}.

A great advantage of these Bayesian methods is their ability to infer dynamics when the system under consideration is not isolated, but is influenced by its environment and other processes to which it may be weakly coupled \cite{Haken:75,Kloeden:11,Suprunenko:13,Clemson:14}. One manifestation of such external influence occurs when the underlying dynamical systems are subject to noise, and it has already been studied in detail theoretically \cite{Stankovski:12b,Duggento:12,Daunizeau:09,Luchinsky:08,Duggento:08,Smelyanskiy:05a}. When dealing with data from natural systems one should allow for the possibility of the dynamics being time-varying. In such cases, the Bayesian method \cite{Stankovski:12b,Duggento:12} published recently is of particular interest: it can identify time-varying dynamics even in the presence of noise, and it is able to follow the time-evolution of the parameters. It is the latter method that provides the main focus of this tutorial.

The method can be applied to various types of dynamical system. In what follows, however, we will focus our attention on coupled oscillatory dynamical systems. The latter frequently arise in physiology and include, for example, the cardio-respiratory or neuronal systems \cite{Stefanovska:00a,Shiogai:10,Rudrauf:06,Jirsa:13} which are, in fact, coupled dynamical systems with time-varying parameters that are subject to external noise. In that context, one might be interested in e.g.\ detecting causal interactions and the directionality of influence \cite{Rosenblum:01,Palus:03a,Porta:14,Jamsek:10}, or in coherence and synchronization \cite{Tass:98,Mormann:00,Stankovski:14a,Rulkov:95}. Of special interest is the recently developed detection and description of interactions in terms of \emph{coupling functions} \cite{Kralemann:13b,Kiss:05,Stankovski:12b}. Dynamical Bayesian inference has already been used to investigate how the cardiorespiratory coupling functions are affected by aging \cite{Iatsenko:13a}. The same method has also been applied in a quite different context, namely, to facilitate the use of  inter-oscillator coupling functions to improve the security of communications systems \cite{Stankovski:14a}.

In the tutorial presented below, we discuss the practicalities of how this particular Bayesian method can be implemented, including the algorithms, programming and applications. The problems and phenomena that were originally treated by the method \cite{Stankovski:12b,Duggento:12,Luchinsky:08,Duggento:08,Smelyanskiy:05a}
were relatively complex and had to be presented in a rather compressed format. Given that the tutorial is intended for a wider audience, we adopt here a different presentation style using examples that are simple and an exposition that is quite detailed. The tutorial is organized as follows. First, we start by summarising the background and basics of Bayesian probability as proposed by Thomas Bayes, the originator of this inference approach. Then in section \ref{s:implementation} we introduce the theoretical terms needed for the implementation of dynamical Bayesian inference. The algorithms and the programming description are given in section \ref{s:Algorithm}.  Applications of the method to three examples are discussed in section \ref{s:Examples}. The first of these uses coupled phase oscillators to present the basics of the inference of time-evolving phase dynamics in the presence of noise. The second example uses coupled limit-cycle oscillators, also describing the reconstruction and detection of synchronization and coupling functions. The third and final example discusses the inference of coupled chaotic systems in state space, as an implementation of the secure encryption technique. Finally, in section \ref{s:conclusion}, we discus possible generalizations of the method, consider the implications for other areas, point out the relationships to other methods, and offer some concluding remarks.

\section{The legacy of Thomas Bayes (1701-1761) }\label{s:legacy}

It was fortunate for Thomas Bayes' legacy that his friend Richard Price significantly edited and updated his work, and read it posthumously to the Royal Society on his behalf in 1763.
It was published in \emph{Philosophical Transactions of the Royal Society of London} the following year. The ideas gained only limited exposure until they were independently rediscovered and further developed by Laplace, who first published their modern formulation in his 1812 \emph{Th\'{e}orie analytique des probabilit\'{e}s}.

The classical approach to statistics defines the probability of an event as ``The number of times the event occurs over the total number of trials, in the limit of an infinite series of equiprobable repetitions''. Many of the limitations inherent in this definition can be avoided, and paradoxes resolved, by taking a Bayesian stance about probabilities. Bayes defines probability as:

\begin{quote}
``The probability of any event is the ratio between the value at which an expectation depending on the happening of the event ought to be computed, and the value of the thing expected upon its happening.''

\end{quote}

\noindent However even Bayes himself might not have embraced the broad interpretation now referred to as Bayesian.  It is difficult to assess Bayes' philosophical views on probability, because his work does not go into questions of interpretation.

Today Bayesian probability is used to describe several different, but related, interpretations of probability. To evaluate the probability of a hypothesis, Bayesian probability specifies some prior probability, which is then updated in the light of new, relevant data. ``Bayesian'' has been used in this sense since the rebirth of Bayes' ideas in the 20th century. Advances in computer technology have allowed scientists from many disciplines to extend the approach and to apply it in diverse fields.  Sir Harold Jeffreys, who wrote the book \emph{Theory of Probability}, which first appeared in 1939, played an important role in the revival of the Bayesian view of probability. He wrote that Bayes' theorem ``is to the theory of probability what Pythagoras's theorem is to geometry''.

So what exactly is Bayes' theorem? Scientific hypotheses are typically expressed through probability distributions for observable data $\mathcal X$ which depend on the model parameters $\mathcal M$. In the Bayesian framework, current knowledge about the model parameters is expressed by placing a probability distribution on the parameters, called the ``prior distribution'', often written as $p_{\mbox{\scriptsize prior}}(\mathcal M)$. When new data $\mathcal X$ become available, the information they contain regarding the model parameters is expressed in the ``likelihood,'' which is proportional to the conditional distribution of the observed data given the model parameters $\ell ( \mathcal X | \mathcal M )$. This information is then combined with the prior to produce an updated probability distribution called the ``posterior distribution,'' on which all Bayesian inference is based. Bayes' theorem, an elementary identity in probability theory, states how the update is done mathematically -- the posterior is proportional to the prior times the likelihood, over the whole available parameter space:

\begin{equation}\label{eq:bayes}
p_{{\mathcal X}}(\mathcal M | \mathcal X) = \frac{\ell ( \mathcal X | \mathcal M ) \,
p_{\mbox{\scriptsize prior}}(\mathcal M) }{ \int{\ell ( \mathcal X | \mathcal M
) \, p_{\mbox{\scriptsize prior}}(\mathcal M) d \mathcal M}}.
\end{equation}

\noindent From this point of view, the task to be faced in Bayesian analysis is to construct and evaluate the likelihood function, given that the data and the prior knowledge are available.

This single line of Bayes' theorem appears simple, but has a rather profound meaning. It gives one a means of reversing the problem of inference -- starting from observations to get back to the nature of the causation. Even better for practical applications is that it works with probabilities, expressing ``beliefs'' or the level of how likely something is to happen. It is a powerful method that provides natural ways for people in many disciplines to structure their data and knowledge, and to yield direct and intuitive answers to their practical questions.

Today, Bayesian theory enjoys wide interpretation and application over most of science and experimental areas quite generally. Since Bayes first propounded his ideas, there have been huge developments in both the theory and applications. The latter span practically every aspect of science, including particle physics, astrophysics, cosmology, geophysics, communications, pharmacology and medical and biological physics \cite{Toussaint:11,Arulampalam:02,Friston:02,Manolopoulou:12,Lemey:09,Smelyanskiy:05a,Stankovski:12b,Stankovski:14a}. There are entire societies, such as the \emph{International Society for Bayesian Analysis} (ISBA), special conferences are organized on Bayesian analysis, and there exist numerous journals specialized in Bayesian theory and analysis. Given the current level of activity in the field, Bayesian inference promises to become even more useful and involved in solving the great scientific and everyday problems faced by humanity.

\section{Dynamical Bayesian inference}\label{s:implementation}
In the present context, dynamical inference refers to a procedure for inferring a model in terms of differential equations based on the analysis of a time-series. The method is based on a development of Feynman's path integral whose central idea is that, for the motion of a particle between two points in space, all possible connecting trajectories should be considered and a probability amplitude assigned to each one of them. This path integral gives the likelihood of observation of a dynamical trajectory for a given set of distributions of the model parameters. Once the actual dynamical trajectory is measured in the experiment, the distributions for the set of model parameters can be improved by use of Bayes' theorem. The other main feature, characteristic of the method, is that the Bayesian framework is applied to a model whose deterministic part is allowed to be time-varying.

The aim is to provide a method that can infer a model of two (or more) weakly-interacting systems subject to noise:
\begin{equation}
\dot \cci_i = \ff(\cci_i,\cci_j|\cc)+ \sqrt{\EE}\xi_i,
\label{eq:model}
\end{equation}
where $i\neq j={1,2}$, and $f(\cci_i,\cci_j|\cc)$ is the deterministic part of the internal and the interacting dynamics. The vector $\cc$ denotes the parameters of the model. The dynamical noise is assumed to be white, Gaussian, and parameterized by a noise diffusion matrix ($\EE$ $2 \times 2$): $\langle \xi_i(t) \xi_j(\tau)\rangle = \delta(t-\tau) E_{ij}$. In what follows we exploit the Bayesian method presented in \cite{Stankovski:12b,Duggento:12,Luchinsky:08,Duggento:08,Smelyanskiy:05a}, and the readers interested in additional theoretical details are directed to those papers and to the references therein.

At this point we speak of $\cci_i$ in general, but later we will refer separately to the phase or state domain, depending on the type of data that we are inferring. When we apply the method to analyze different systems, the base functions are the only thing that will change in the inferential framework. Here, we will use polynomial base functions for the state domain, while for the phase domain we will decompose the dynamics into Fourier components:
\begin{equation}
\begin{split}
\dot \pphi_i=& \sum_{k=-K}^{K} c^{(i)}_k \, \Pphi_{i,k}(\pphi_i,\pphi_j)  + \sqrt{\EE} \xxi_i,
\label{eq:phiF}
\end{split}
\end{equation}
where $\Pphi_{1,0}=\Pphi_{2,0}=1$, $c^{(i)}_0=\omega_i$ are the respective frequencies, and the rest of the $\Pphi_{i,k}$ and $c^{(i)}_k$ are the $K$ most important Fourier components serving as base functions.

Given that $2 \times N$ time-series ${\mathcal X} = \lbrace {\bf \cci}_{n} \equiv \cci(t_{n}) \rbrace$ ($t_n=nh$) are provided, and assuming that the model base functions are known, the main task for dynamical Bayesian inference \cite{Smelyanskiy:05a,Luchinsky:08} is to infer the unknown model parameters and the noise diffusion matrix ${\mathcal M}=\{ \cc,\EE \}$. The problem eventually reduces to maximization of the conditional probability of observing the parameters ${\mathcal M}$, given the data ${\mathcal X}$. For this we applied Bayes' theorem, given before as (\ref{eq:bayes}).
The prior distribution, enclosing previous knowledge of the unknown parameters based on observations, is assumed to be known. The task is therefore to determine the likelihood functions in order to infer the final posterior result. If the sampling $h$ is small enough, using the acquired time-series one can construct the Euler midpoint approximation of Eqs.\ (\ref{eq:model}):
\begin{equation}
 \cci_{i,n+1} =\cci_{i,n}+ h \ff(\cci_{i,n}^*,\cci_{j,n}^*|\cc)+ h \sqrt{\EE}\zz_n,
\label{eq:discr}
\end{equation}
where $\cci_{n}^*=(\cci_{n+1}+\cci_{n})/2$ and $\zz_n$ is the stochastic integral of the noise term over time:
$\zz_n \equiv \int_{t_n}^{t_{n+1}} \zz(t) \, dt = \sqrt{h}\,\Hh\, {\xxi_n} \,
$ for the $\Hh$ matrix that satisfies the Cholesky decomposition $\Hh \Hh^ \TT = \EE$. The parameters $\cc$ act as a scale coefficients for the base functions $\ff(\cci_i,\cci_j|\cc)=\cc \mathbf P(\cci_i,\cci_j)$.
Use of the stochastic integral for noise that is white and independent leads to a likelihood function that is given by the product over $n$ of the probability of observing $\cci_{n+1}$ at each
time. The negative log-likelihood function is then $S=-\ln \ell({\mathcal X}|{\mathcal
M})$ given as:
\begin{equation}
\begin{split}
    S &=   \frac{N}{2}\ln |{\EE}| + \frac{h}{2}\, \sum_{n=0}^{N-1}\Big(
     \cc \frac{\partial \mathbf P(\cci_{\cdot,n}) }{\partial \cci}+\\
     &+ [\dot{\cci}_{n} - \cc \mathbf P({\cci}_{\cdot,n}^{\ast})]^T {({\EE}^{-1})}  [\dot{\cci}_{n} - \cc \mathbf P({\cci}_{\cdot,n}^{\ast})] \Big ),
\end{split}
    \label{eq:likelihood}
\end{equation}
where $\dot \cci_{n}=(\cci_{n+1}-\cci_{n})/h$ and the dot index in ${\cci}_{\cdot,n}$ represents the appropriate ($i$ or $j$ in this case) index. The likelihood (\ref{eq:likelihood}) is of quadratic form and, if the prior is a multivariate normal distribution, so also will be the posterior. Given such a distribution as a prior for the parameters $\cc$, with mean $\bar {\cc}$, and covariance matrix ${ {\bf \Sigma_{\mbox{\scriptsize prior}} \equiv \Xi}^{-1}}_{\mbox{\scriptsize prior}}$, the final stationary point of $S$ is calculated recursively from the following four equations:\\
\begin{itemize}
\item[(a)] the noise matrix $\EE$
\begin{equation}
    \label{eq:inf1_E}
     \EE  = \frac{h}{N} \left(
 \dot{\cci} - \cc\mathbf P({\cci}_{\cdot,n}^{\ast}) \right)\left(\dot{\cci}_{n} - \cc\mathbf P({\cci}_{\cdot,n}^{\ast}) \right)^T ,
\end{equation}
\item[(b)] the concentration matrix $\mathbf \Xi$
\begin{equation}\label{eq:inf2_Xi}
     {\mathbf \Xi}  = {{\mathbf \Xi}_{\text{prior}}}   + h \, \mathbf P({\cci}_{\cdot,n}^{\ast})^T \,
{(\EE^{-1})} \,   \mathbf P({\cci}_{\cdot,n}^{\ast}),
\end{equation}
\item[(c)] a temporary matrix variable $\bf r$
 \begin{equation}\label{eq:inf3_r}
     {\mathbf r}   = {\mathbf \Xi}_\text{prior} \,  {\cc} + h \, \mathbf P({\cci}_{\cdot,n}^{\ast}) \,
(\EE^{-1}) \, \dot{{\cci}}_{n} - \frac{h}{2} \mathbf v ({\cci}_{\cdot,n}^{\ast}),
 \end{equation}
where the components of matrix $\mathbf v$ are the partial derivatives of the base functions\\
 $ \mathbf v ({\cci}_{\cdot,n}^{\ast})=\frac{\partial \mathbf P(\cci_{\cdot,n}^{\ast}) }{\partial \cci_{\cdot}}$,\\
\item[(d)] and the final parameters $\cc$
\begin{equation}\label{eq:inf4_c}
     \cc = {\mathbf \Xi}^{-1} {\mathbf r} , \\
\end{equation}
\end{itemize}
where summation over $n=1,\ldots,N$ is assumed. The first initial prior can be set as a non-informative flat normal distribution, ${{\bf\Xi}}_{\text{prior}}=0$ and $\bar \cc_{\mbox{\scriptsize prior}}=0$.

By evaluating the four equations (\ref{eq:inf1_E})-(\ref{eq:inf4_c}) using the readout time series ${\mathcal X}$, one can calculate effectively the multivariate probability ${\mathcal N}_{\mathcal X}(\cc|,\bar{\cc},\Xxi)$ which explicitly defines the probability density of each parameter set of the dynamical system.

The inference method needs to follow the time-evolution of the parameter set $\cc$ while separating dynamical effects from the noise. In order to achieve this, we modify the propagation procedure between the covariance of the current posterior $\bf \Sigma_{\text{post}}^n$  and the next prior $\bf \Sigma_{\text{prior}}^{n+1}$ \cite{Stankovski:12b}. The definite matrix $\bf \Sigma_{\text{diff}}$ is introduced in order to show how much each parameter diffuses normally. Thus, the next prior probability of the parameters is the convolution of two current normal multivariate distributions, $\bf \Sigma_{\text{post}}$ and $ \bf \Sigma_{\text{diff}}$: $\bf \Sigma_{\text{prior}}^{n+1} =
\bf \Sigma_{\text{post}}^n + \bf \Sigma_{\text{diff}}^n$. To avoid propagation in the assumptions about correlation between parameters, we consider $\bf \Sigma_{\text{diff}}$ to be diagonal ($\rho_{ij}=\delta_{ij}$ in \cite{Stankovski:12b}). We assume each standard deviation $\sigma_i$ to be a known fraction of the relevant standard deviation from the posterior covariance (or parameters)  $\sigma_i = p_w (\sigma^n_{\text{post}})_i$, where $p_w$ is a constant parameter. In practice this means that $\bf \Sigma_{\text{diff}}$ has zero values everywhere, except for the diagonal values, which are a fraction of the diagonal values of the posterior $\bf \Sigma_{\text{post}}$.

\section{Algorithms and programming}
\label{s:Algorithm}
In this section we discuss the algorithmic and programming details needed for the implementation. We start by presenting what is arguably the most complicated part -- the algorithm for dynamical Bayesian inference applied within a single window of readout data. The algorithm employing recursion using the Eqs.\ (\ref{eq:inf1_E})-(\ref{eq:inf4_c}) can be summarized in terms of the following steps:
\begin{itemize}
\item[i)] the algorithm starts from a  $ \cc_{\text{prior}}$ and $\mathbf \Xi_\text{prior}$,
\item[ii)] noise matrix  $\mathbf E_\text{new}$ is calculated using Eq.\ (\ref{eq:inf1_E}),
\item[iii)] $\mathbf \Xi_\text{new}$ is calculated using Eq.\ (\ref{eq:inf2_Xi}),
\item[iv)] $\mathbf r$ is calculated using Eq.\ (\ref{eq:inf3_r}),
\item[v)] $\mathbf c_\text{new}$ is calculated using Eq.\ (\ref{eq:inf4_c}),
\item[vi)] then again to point ii) using $\mathbf c_\text{new}$ as $\mathbf c$.
\end{itemize}
The stopping rule is that ``convergence'' has been reached i.e.\ when further iteration of the algorithm would not modify $\mathbf c$ and $\mathbf \Xi$ any more. For example, we used  the condition: $\sum{(\mathbf c_{old}- \mathbf c_{new})^2/\mathbf c_{new}^2 <\eps}$ where $\eps$ is some very small constant. Because the problem is parabolic, this convergence is very fast -- typically a few cycles.
The initial prior distribution is assumed to be a noninformative ``flat'' distribution, representing the initial limit of an infinitely large normal distribution, and obtained by setting ${{\bf\Xi}}_{\text{prior}}=0$ and $\cc_{\mbox{\scriptsize prior}}=0$.

For general programming purposes, we now outline an informal pseudo-code description of the main algorithms and sub-algorithms. Comments are presented in grey. First we describe the algorithm for Bayesian inference:\\
\line(1,0){250}\\
Algorithm 1: \textbf{Bayesian inference}\\
\\ \gray{$\backslash \backslash$calculate temporary variables beforehand}\\
$\text{ }\text{ }\text{ }$ -- calculate \textbf{P}\\
$\text{ }\text{ }\text{ }$ -- calculate \textbf{v}\\
\\$\cc_\text{pt}=\cc_\text{pr}$\\
FOR lp=1:MaxLoops $\text{ }\text{ }\text{ }$ \gray{$\backslash \backslash$main recursive loop}\\
$\text{ }\text{ }\text{ }$ -- calculate $\mathbf E$\newline
$\text{ }\text{ }\text{ }$ -- calculate $\cc_\text{pt}$\\
$\text{ }\text{ }\text{ }$ IF SUM($(\cc_\text{pr}-\cc_\text{pt})^2/\cc_\text{pt}^2$)<SmallError\\
$\text{ }\text{ }\text{ }$$\text{ }\text{ }\text{ }$ RETURN\\
$\text{ }\text{ }\text{ }$ ENDIF\\
$\text{ }\text{ }\text{ }$ $\cc_\text{pr}=\cc_\text{pt}$\\
ENDFOR\\
\line(1,0){250}\\
\\ The sub-algorithms `\emph{calculate P}' and `\emph{calculate v}' depend on the particular base functions and their partial derivatives. These two functions depend on the specific model to be inferred, and they are the only part needing change if one selects a different model for the inference. The sub-algorithms only involve the simple evaluation of base functions in respect of the ${\cci}^{\ast}$ time series, and will not be discussed in detail. The other main calculations are performed within `\emph{calculate} $\mathbf E$' and `\emph{calculate} $\cc_\text{pt}$', which are discussed in detail below.\\
\line(1,0){250}\\
Algorithm 2: \textbf{calculate} $\mathbf E$ $\text{ }\text{ }\text{ }$ \gray{$\backslash \backslash$use of Eq.\ (\ref{eq:inf1_E})}\\
\\ $\EE$ = $\EE$+($\dot \cci$-\textbf{P}*$\cc$)*(transpose of ($\dot \cci$-\textbf{P}*$\cc$)) \\
$\EE$ = h/N*$\EE$\\
\line(1,0){250}\\

\noindent Finally, the algorithm for calculation of the parameters used in the main recursive loop is expressed as: \\
\line(1,0){250}\\
Algorithm 3: \textbf{calculate} $\cc$\\
$\mathbf{invE}$ = inverse of $\mathbf E$\\
\\ \gray{$\backslash \backslash$calculate $\Xxi$, Eq.\ (\ref{eq:inf2_Xi})}\\
FOR i=1:$l$\\
$\text{    }$ FOR j=1:$l$ \\
$\text{    }$ $\text{     }$ $\Xxi_\text{pt}((i-1) \cdot K+1 \text{  TO } i \cdot K, (j-1) \cdot K+1 \text{  TO } j \cdot K)$=$\\
\text{    } \text{    } \text{    } \text{    }\Xxi_\text{pr}((i-1) \cdot K+1 \text{  TO } i \cdot K, (j-1) \cdot K+1 \text{  TO } j \cdot K)$+h*$\mathbf{invE}(i,j)$*\textbf{P}*(transpose of \textbf{P})\\
$\text{    }$ ENDFOR\\
ENDFOR\\
\\ \gray{$\backslash \backslash$calculate $\rr$, Eq.\ (\ref{eq:inf3_r})}\\
$\mathbf{ED}$=$\mathbf{invE}$*$\dot \cci$\\
FOR i=1:$l$\\
$\text{    }$ FOR j=1:$l$ \\
$\text{    }$ $\text{     }$  $\rr (\text{ALL},i)$=$\rr (\text{ALL},i)$+$\Xxi_\text{pr}((i-1) \cdot K+1 \text{  TO } i \cdot K, (j-1) \cdot K+1 \text{  TO } j \cdot K)$*$\cc(\text{ALL},j)$
$\text{    }$ ENDFOR\\
$\text{    }$ $\rr(\text{ALL},i)$=$\rr (\text{ALL},i)$+h $\cdot$ \textbf{P}*(transpose of $\mathbf{ED}$)-h/2$\cdot$ sum(\textbf{v}$(\text{ALL},i)$)\\
ENDFOR\\
\\ \gray{$\backslash \backslash$calculate $\cc$, Eq.\ (\ref{eq:inf4_c})}\\
$\cc$=(inverse of $\Xxi_\text{pt}$)*$\rr$\\
\line(1,0){250}\\
where $l$ is the number of time-series provided, and $K=M/l$ with $M$ being the total number of base function used. For matrix calculations, one should interpret \emph{TO} as the linear span of integer indices within a column or row of a matrix, and similarly \emph{ALL} as all the respective indices within a column or row of a matrix. Note that Eqs.\ (\ref{eq:inf2_Xi}) and (\ref{eq:inf3_r}) are implemented with $l=2$ \emph{for}-cycles here, but for general $l$ one should include $l$ nested \emph{for}-cycles.

Given in this vectorized form, the algorithms are quite efficient and the processing time needed is very short. In the implementation one should simply follow the dimensions of the vectors and matrices to reach the correct evaluation. For example, if we are given two ($l=2$) time-series of length $N=4000$ and we use six base functions in total ($M=6$) with three ($K=M/l=3$) base functions for each data time series, then the respective dimensions of the matrices are: $\EE_{l \times l}=\EE_{2 \times 2}$, $\rr_{K \times l}=\rr_{3 \times 2}$, $\Xxi_{M \times M}=\Xxi_{6 \times 6}$, $\cc_{k \times l}=\cc_{3 \times 2}$, $\PP_{K \times N}=\PP_{2 \times 4000}$ and $\vv_{l \times K \times N}=\vv_{2 \times 3 \times 4000}$. For matrix implementation, it makes sense for some of the variables to be broken into parts like $\vv_{l \times K \times N}=\{\vv1_{ K \times N},\vv2_{ K \times N}\}$, or to be filled in parts e.g.\ $\Xxi_{M \times M}$ can be filled in four $\Xxi_{K \times K}$ parts. Note that in this way the dimensional representation of the vectors is slightly altered from the theoretical notation in order to reach the vectorized form in the interests of faster calculations.

The three algorithms described are applied to a single window of data. The time-series are separated into sequential blocks, and the algorithms are applied to each of them in turn. The core of dynamical Bayesian inference is that it uses informative priors i.e.\ the evaluation of the next block of data depends on and uses the evaluation results from the previous block. The process of information propagation, between the $n$ posterior and the next $n+1$ prior distribution, can be adjusted to allow the time-variability of the parameters to be followed. We used propagation depending on the concentration matrix $\Xxi_\text{pt}$ or on the vector of parameters $\cc_\text{pt}$. We now describe propagation with respect to the concentration matrix $\Xxi_\text{pt}$:\\
\line(1,0){250}\\
Algorithm 4: \textbf{Propagation}\\
$\cc_\text{pr}^{n+1}=\cc_\text{pt}^{n}$\\
\\$\mathbf{inv}\Xxi^n$ = inverse of $\Xxi_\text{pt}^n$;$ \\
\mathbf{invDiff}^n$=0 \\
FOR i=1:K  \\
$\text{ }\text{ }\text{ }$ $\mathbf{invDiff}^n$(i,i)=$p_w^2$*$\mathbf{inv}\Xxi^n$(i,i)\\
ENDFOR\\
$\Xxi_\text{pt}^{n+1}$=inverse of ($\mathbf{invDiff}^n$+$\mathbf{inv}\Xxi^n$)\\
\line(1,0){250}\\

Once the inference has been performed, one can use the inferred parameters to detect certain dynamical and phenomenological characteristics of the interacting systems. For example, calculating the norm of the inferred parameters together with the relevant base functions, one can detect the coupling strength and directionality between the oscillators \cite{Rosenblum:01,Palus:03a,Jamsek:10,Staniek:08,bahraminasab:08}. Similarly, one can reconstruct the form of the coupling function \cite{Stankovski:12b,Kralemann:07,Tokuda:07,Miyazaki:06,Kiss:05}. One can determine whether or not the systems are synchronized \cite{Rosenblum:96,Rulkov:95,Kocarev:96}, once the parameters have been inferred. For the case of phase synchronization this can be done, for example, with a map representation and the modified Newton root-finding method (see \cite{Duggento:12} for details), while for generalized synchronization it can be done by evaluation of the largest Lyapunov exponents \cite{Stankovski:14a}. In this way both phase and generalized synchronization can be detected within the same framework based on the dynamical Bayesian inference presented. 
It is worth pointing that, by adopting this approach, one detects intrinsic synchronization based on the effective connectivity \cite{Park:13}, which differs from other synchronization detection methods based on statistics of the phase and state time-series \cite{Tass:98,Mormann:00,Rulkov:95,Parlitz:96}. This can sometimes be advantageous.
\begin{figure}[b!]
\begin{center}
\includegraphics[width=0.9\linewidth,angle=0]{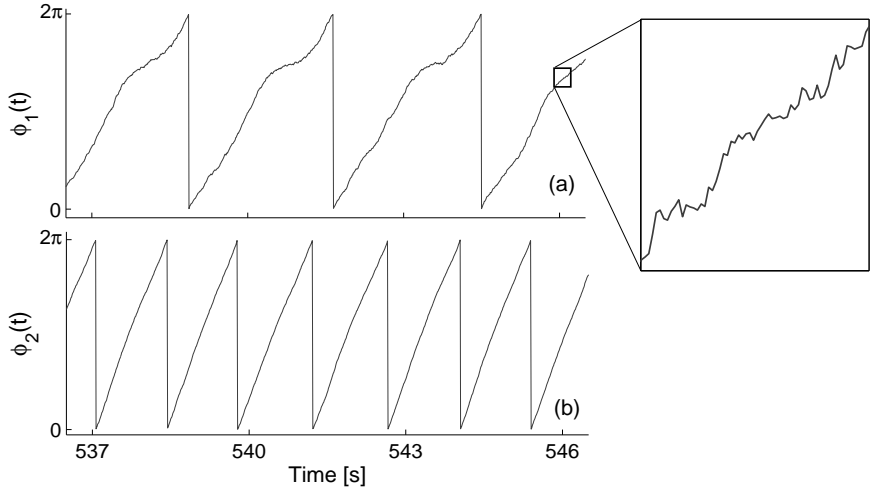}
\caption{\label{fig1:phs} The instantaneous phases generated by the model (\ref{eq:phs}) of two coupled phase oscillators. (a) The phase $\phi_1$ and (b) $\phi_2$. For simpler presentation, the phases are ``wrapped'' within $2\pi$. The enlarged panel on the right illustrates the noise perturbations of $\phi_1$ on an expanded scale.}
\end{center}
\end{figure}

\section{Examples}
\label{s:Examples}

In this section we outline three examples of the reconstruction of coupled oscillators that have time-varying dynamics and which are also subject to noise. The first example illustrates the basics of the inference method on a simple phase oscillator model, while the second example involves limit-cycle oscillators and presents the detection of several characteristics and relationships that can be detected from the inferred parameters. The third example presents the inference of coupled chaotic systems in state space. MatLab codes for the examples are provided at the following link\footnote{http://py-biomedical.lancaster.ac.uk/}.

\begin{table*}  \label{tab:par}
\begin{center}
\begin{tabular}{c c  c  c  c  c  c c c c c}
\hline
   Parameters  & \hspace{0.2cm} $\omega_1$ \hspace{0.25cm}\vline & \hspace{-0.1cm}$\omega_2$ \hspace{0.25cm}\vline & $a_1$ \hspace{0.25cm}\hspace{0.0cm}\vline & \hspace{0.1cm}$a_2$ \hspace{0.25cm}\hspace{0.0cm}\vline &\hspace{0.2cm} $a_3$ \hspace{0.3cm}\vline & \hspace{0.cm}$a_4$\hspace{0.35cm} \vline & $E_{11}$ \vline & \hspace{-0.4cm} $E_{12},E_{21}$ \vline &\hspace{-0.2cm} $E_{22}$ \\

\hline \hline
Intrinsic values   &2.032& \hspace{-0.2cm}4.53 & \hspace{-0.25cm}0.8 & \hspace{-0.2cm}0 & \hspace{-0.2cm}1.013 &\hspace{-0.5cm} 0.6& \hspace{-0.2cm}0.03&\hspace{-0.4cm}0& \hspace{-0.2cm}0.01\\
Inferred means     &2.026&\hspace{-0.2cm}4.537&\hspace{-0.25cm}0.803&\hspace{-0.2cm}-0.014&\hspace{-0.2cm}1.054&\hspace{-0.5cm}0.596 & \hspace{-0.2cm} 0.029 &\hspace{-0.4cm}0.000&\hspace{-0.2cm}0.010 \\
\hline
\end{tabular}
\end{center}
\caption{Results from the inference of the numerically simulated system (\ref{eq:phs}). The
first row describes the physical meaning of the parameters, and the second and third two rows show, respectively, the actual values of the parameters and their inferred mean values. The results are presented for one window of data around $t=1980$ s.}
\end{table*}

\subsection{Coupled phase oscillators}

In order to present in a transparent way the basics of the inference technique we first consider two coupled phase oscillators \cite{Kuramoto:84} subject to noise:
\begin{equation}
\begin{split}
\dot{\phi_1}&=\omega_1(t)  +a_1 \sin(\phi_1)+a_3(t) \sin(\phi_2) + \xi_1(t)\\
\dot{\phi_2}&=\omega_2  +a_2 \sin(\phi_1)+a_4 \sin(\phi_2)+ \xi_2(t).
\end{split}
\label{eq:phs}
\end{equation}

\begin{figure}[b!]
\begin{center}
\includegraphics[width=0.93\linewidth,angle=0]{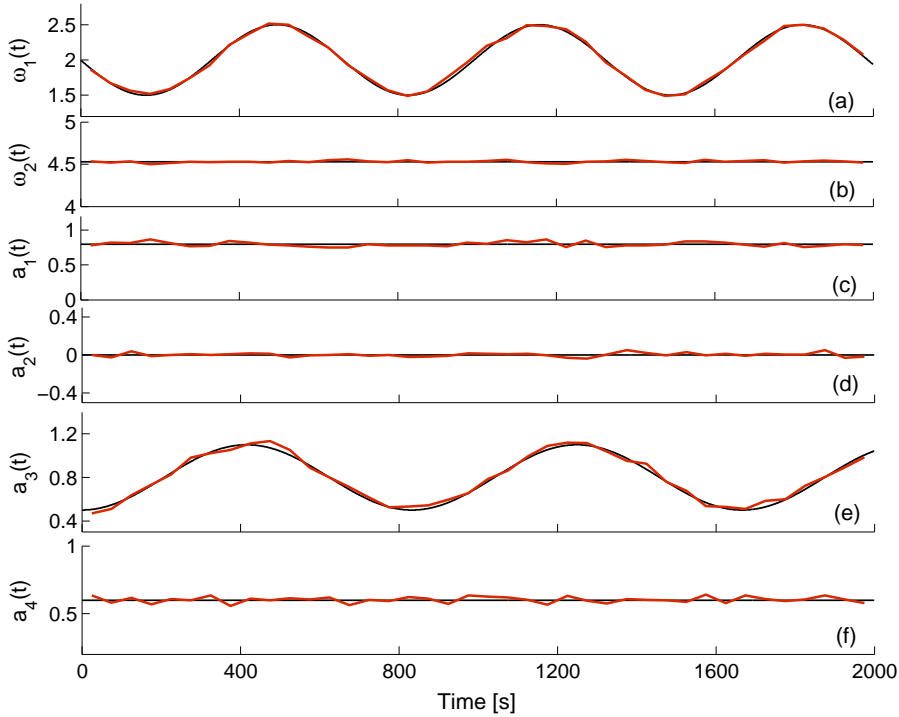}
\caption{\label{fig2:para} Time-evolution of the parameters inferred from model (\ref{eq:phs}). (a) and (b) present the two frequencies, (c) and (f) are the self-dynamics parameters, and (d) and (e) present the coupling parameters. In each case, the actual values of the parameters are indicated by the black curves underlying their inferred values (red).}
\end{center}
\end{figure}

Each oscillator is described by the frequency parameters $\omega_1$, $\omega_2$, the parameters for their self-dynamics $a_1$, $a_4$ and the coupling parameters $a_2$, $a_3$ for the direct influence coming from the other oscillator. Two parameters are set to be periodically time-varying, the frequency  $\omega_1(t)=2-0.5\sin(2\pi0.00151 t)$ and the coupling parameter $a_3(t)=0.8-0.3\sin(2\pi0.0012 t)$. The noises are set to be white Gaussian and mutually uncorrelated. The other parameter values are $\omega_2=4.53$, $a_1=0.8$, $a_2=0$, $a_4=0.6$, $E_{11}=0.03$ and $E_{22}=0.01$. Fig.\ \ref{fig1:phs} shows samples from the resultant time series to which dynamical Bayesian inference is to be applied.

The choice of phase oscillators is very convenient for inference, because one needs to reconstruct the phase dynamics. Therefore, the phase model is known beforehand and the deterministic terms of the \emph{rhs} of the coupled system (\ref{eq:phs}) are the actual base functions to be used for inference of the six parameters ($\omega_1$, $\omega_2$, $a_1$, $a_2$, $a_3$ and $a_4$). Inference results from a single block of data are presented in Table \ref{tab:par}. The agreement between the actual (intrinsic) parameters and their inferred values is excellent, and the method evidently works to high precision. Additionally -- and which is unique for this method -- the intensity and the correlations of the noise are inferred very precisely. Fig.\ \ref{fig2:para} shows the time-variations of all the parameters inferred from sequential windows of length $w=40$ s, with a propagation constant $p_w=0.2$. It is clearly evident that the parameters and their time-variability are inferred precisely.

\subsection{Coupled limit-cycle oscillators -- phase domain inference}

\begin{figure}[b!]
\begin{center}
\includegraphics[width=0.78\linewidth,angle=0]{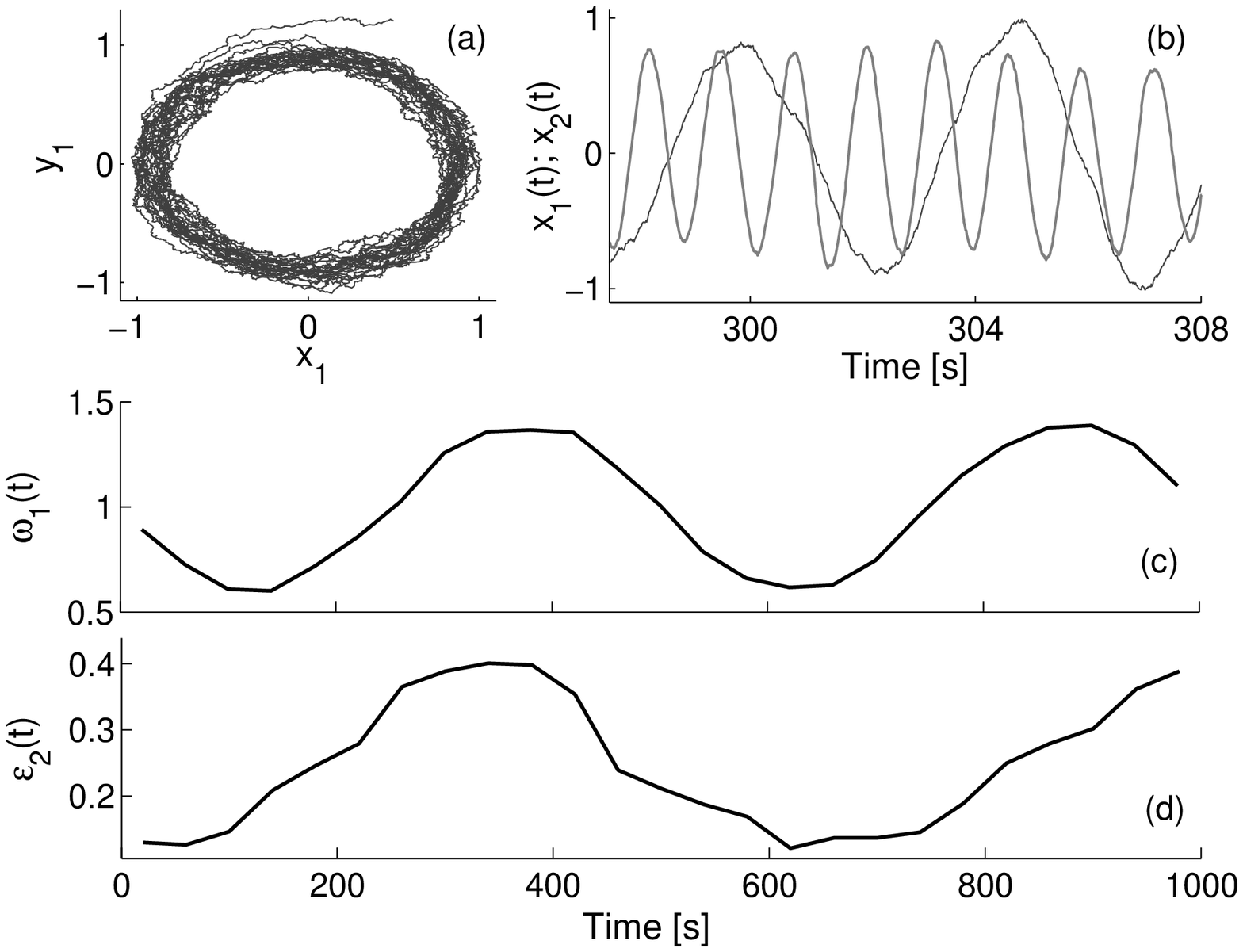}
\caption{\label{fig3:LCpara} Inference of time-varying parameters from the model (\ref{equ:num_model}) of two coupled limit-cycle oscillators.  (a) Phase portrait  showing the noisy state of the first oscillator. (b) The time-series of the two oscillators $x_1(t)$ and $x_2(t)$. (c) The inferred time-evolution of the frequency parameter, and (d) the inferred net coupling from the first to the second oscillator).}
\end{center}
\end{figure}

The second example involves a system of two coupled limit-cycle oscillators, which can serve as a model for a number of oscillatory processes that occur in nature, including electrochemical, mechanical, cardio-respiratory, and other biological systems \cite{Stankovski:12b,Kiss:05,Kralemann:08,Pikovsky:01}. The model consists of two interacting Poincar\'{e} oscillators subject to noise:

\begin{equation}
\begin{split}
\dot x_1&= -  \Big(\sqrt{x_1^2+y_1^2}-1 \Big) x_1  -\omega_1(t) y_1  + \varepsilon_{1} (x_2-x_1)+\xi_1(t)\\
\dot y_1&= - \Big(\sqrt{x_1^2+y_1^2}-1 \Big) y_1  +\omega_1(t) x_1  + \varepsilon_{1} (y_2-y_1)+\xi_2(t)\\
\\
\dot x_2&= -  \Big(\sqrt{x_2^2+y_2^2}-1 \Big) x_2  -\omega_2 y_2  + \varepsilon_{2}(t) (x_1-x_2)+\xi_3(t)\\
\dot y_2&= - \Big(\sqrt{x_2^2+y_2^2}-1 \Big) y_2  +\omega_2 x_2  + \varepsilon_{2}(t) (y_1-y_2)+\xi_4(t),\\
\end{split}
\label{equ:num_model}
\end{equation}
where periodic time-variability is introduced in the frequency of the first oscillator $\omega_1(t)=1-0.4\sin(2\pi0.002 t)$ and in the coupling parameter from the first to the second oscillator $\varepsilon_2(t)=0.2-0.1\sin(2\pi0.0017 t)$. The noises are again white and Gaussian, with no correlations between them. The other parameters are $\omega_2=4.91$, $\varepsilon_1=0.05$, $E_{11}=E_{22}=0.007$ and $E_{33}=E_{44}=0.004$. The systems are simulated in state space, and the corresponding signals and phase portrait of the first system are given in Fig.\ \ref{fig3:LCpara}(a) and (b).

\begin{SCfigure*}
\centering
\includegraphics[width=0.67\linewidth,angle=0]{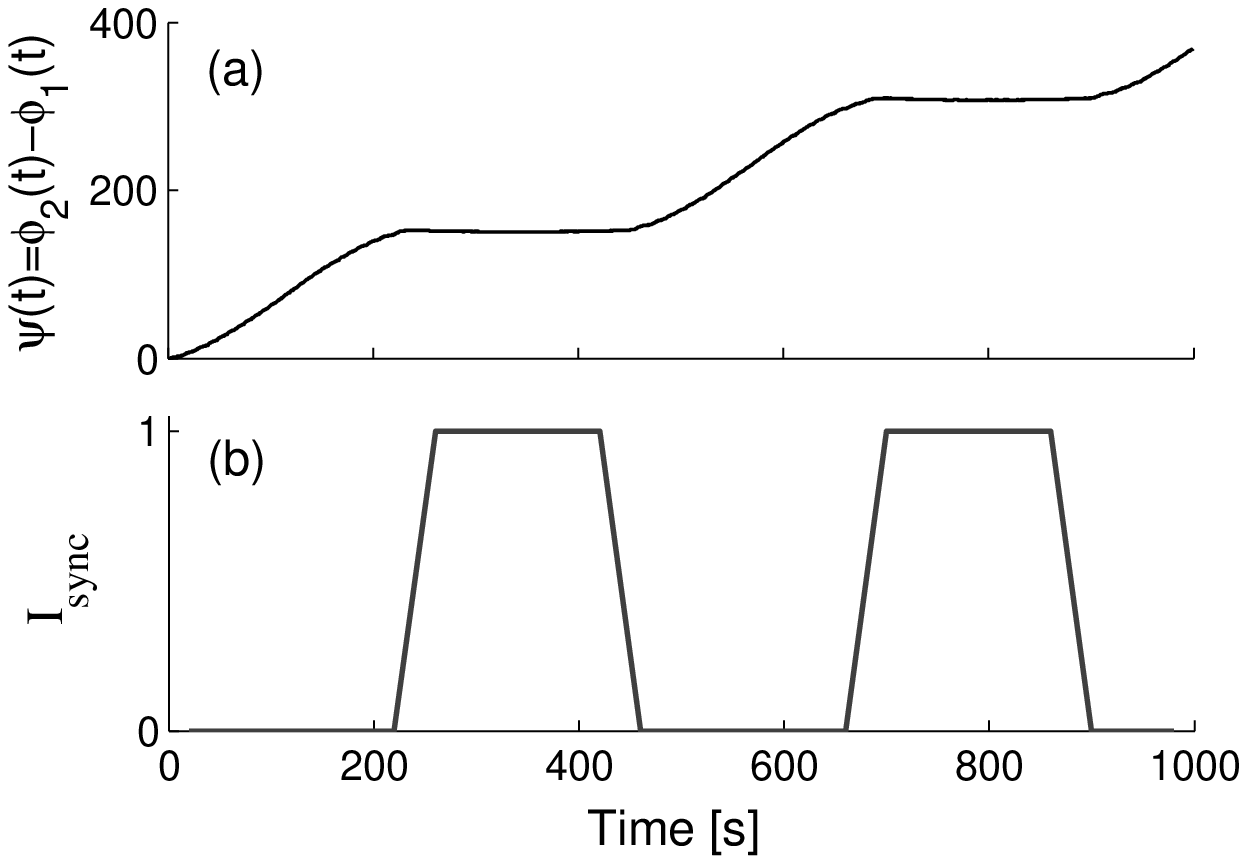}
\caption{\label{fig4:sync} Detection of intermittent synchronization in the system (\ref{equ:num_model}). (a) The phase difference $\psi(t)$ showing the qualitative statistical occurrence of synchronized intervals, which appear as bounded plateaus. (b) The synchronization index demonstrating the detected intrinsic synchronization intervals, evaluated from the inferred parameters. The high values $I_\text{sync}=1$ denote synchronization.  $\text{     }$ $\text{     }$ $\text{     }$ $\text{     }$ $\text{     }$ $\text{     }$ $\text{     }$ $\text{     }$ $\text{     }$$\text{     }$ $\text{     }$ $\text{     }$ $\text{     }$ $\text{     }$ $\text{     }$ $\text{     }$ $\text{     }$ $\text{     }$$\text{     }$ $\text{     }$ $\text{     }$ $\text{     }$ $\text{     }$ $\text{     }$ $\text{     }$ $\text{     }$ $\text{     }$}
\end{SCfigure*}

The phases can be estimated as $\phi_i = \arctan(y_i/x_i)$ ($\arctan$ being a four-quadrant function) from the state signals. Alternatively, one can use the Hilbert \cite{Pikovsky:01} or synchrosqueezed transforms \cite{Daubechies:11}. The choice of base functions for the inference needs to be determined in such a way that the phase dynamics can be reconstructed effectively. Because of their oscillatory nature and periodic solutions, we decompose the phase dynamics into Fourier series. Hence Fourier series up to some order serve as base functions for the dynamical Bayesian inference. However, care must be taken to ensure that none of the functions have strong linear dependences on each other, as this can lead to imprecise and wrong separation of parameters within the inference: for example, choice of $\sin(x)$ and $\sin(-x)$, would rise to problems of precisely this kind because of their mutual linear dependence. In the example, we used only one side of the expansion, e.g.\ for $\sin(n\phi_1+m\phi_2)$ we used the components $n=1,\ldots, K$ instead of $n=-K,\ldots, K$. The reconstruction results for the time-varying parameters are presented in Fig.\ \ref{fig3:LCpara}(c) and (d). The periodic sine variations are evident both in the frequency and the coupling strength. Note that the coupling amplitude is evaluated as the norm of all the relevant inferred parameters that describe this influence.

One can use the inferred parameters, not only to evaluate the characteristics of individual oscillators, but also to determine whether the coupled system undergoes any qualitative transitions.  An obvious example of the latter is the onset or disappearance of synchronization. For this reason we modify the parameters of model (\ref{equ:num_model}) with $\omega_1(t)=1-0.4\sin(2\pi0.0022 t)$, constant coupling parameter $\varepsilon_2=0.2$ and $\omega_2=1.4$. Such a combination of parameters takes the coupled system into and out of synchrony intermittently. The phase difference shown in Fig.\ \ref{fig4:sync}(a) is bounded during the synchronized intervals \cite{Pikovsky:01}. Applying the procedure of return maps and the modified Newton root-finding method \cite{Duggento:12} we can identify the synchronization intervals (Fig.\ \ref{fig4:sync}(b)), which correspond to the intervals of constant phase difference as expected. The map procedure is equivalent to a determination of whether or not the coupled phase oscillator model, with the inferred parameter values, is synchronized. Note also that, during the synchronized intervals, the phases remain almost identical and they do not span enough of the available space for inference (the phase difference appears as non-zero plateaus in Fig.\ \ref{fig4:sync}(a) because it includes an offset corresponding to the phase difference that existed at the start of the synchronization interval considered). This can result in inferred parameters that are far from their intrinsic values. However, the whole set of parameters is again correlated as if it was coming from a synchronized system. One might ask: why do we need to undertake such a complicated procedure to detect synchronization, when something as simple as the phase difference can give a similar answer? The point is that, with the use of the intrinsic inferred parameters, one can distinguish whether or not the phase slips and synchronization transitions are noise-induced \cite{Stankovski:12b}.

\begin{figure}[t!]
\begin{center}
\includegraphics[width=0.98\linewidth,angle=0]{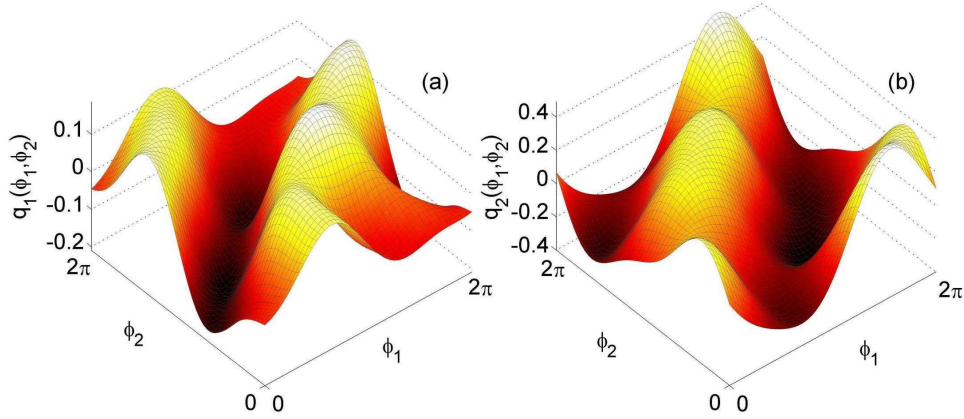}
\caption{\label{fig5:CF} Coupling functions inferred for from the model (\ref{equ:num_model}) of two interacting limit-cycle oscillators. (a) The functional influence $q_1(\phi_1,\phi_2)$ from the second to the first oscillator. (b) The coupling function $q_1(\phi_1,\phi_2)$  from the first to the second oscillator. The functions are inferred within one window of data.}
\end{center}
\end{figure}

Coupling functions are arguably the most important part of the description of the inter-oscillator interactions. They can describe the functional relationship, the law governing the mutual interactions, and the routes to qualitative transitions. As already mentioned above, coupling functions can be used quite generally to describe different aspects of the interactions that occur between a great diversity of oscillatory systems, whether e.g.\ cardio-respiratory, electrochemical or mechanical \cite{Stankovski:12b,Galan:05,Kiss:05,Miyazaki:06,Kralemann:07,Tokuda:07,Pereira:14}. By representation on a $2\pi$-phase grid evaluated for the relevant inferred parameters, we can determine and visualize the phase dynamics very effectively. The coupling functions of systems (\ref{equ:num_model}) inferred from a single window are shown in Fig.\ \ref{fig5:CF}. By inferring the dynamics in a succession of windows, we can follow the time-evolution of the functions \cite{Stankovski:12b}. Moreover, the coupling function $q_i(\phi_i,\phi_j)$ can be further decomposed into self, direct and indirect coupling influences \cite{Iatsenko:13a}, and each of these can be studied separately.

\subsection{Coupled chaotic systems -- state domain inference}

Up to this point, the review has focused mainly on the inference of phase dynamics. The main reason was to present a method that will be quite generally applicable to coupled oscillatory systems. However, there are some situations where the dynamics needs to be analyzed directly from the measured signals in state space. For example, the estimation of phases from chaotic systems can be problematic, while inference in the state domain is directly accessible. If state signals are to be analyzed, then the model equation (\ref{eq:model}) will still hold; the base functions can have e.g.\ a polynomial form, and all the rest of the equations and algorithms can equally be applied to the state signals. For other applications of the inference of (coupled) dynamics in state space see \cite{Stankovski:14a,Smelyanskiy:05a,Luchinsky:08,Duggento:08,Smelyanskiy:03}.

Coupled state space systems, especially chaotic systems, have played an important role in applications to the secure encryption of communications \cite{Cuomo:93,Kocarev:95,Argyris:05,Alvarez:06b}. Recently, a new class of secure communication that is highly resistent to conventual attacks, was introduced \cite{Stankovski:14a} using the same Bayesian method presented in this tutorial. The scheme makes use of the coupling functions between interacting dynamical systems. The information signals are encrypted as the time-variations of independent coupling functions between the coupled systems. Using predetermined forms of coupling function, we can apply dynamical Bayesian inference on the receiver side to detect and separate the information signals while simultaneously eliminating the effect of external noise. The procedure results in an unbounded number of encryption key possibilities, allows the transmission/reception of more than one signal simultaneously, and is robust against external noise. The use of chaotic systems is not essential for the encryption. It does, however, bring an additional level of security associated with the scrambled random-like appearance of chaotic signals.

\begin{figure}[b!]
\begin{center}
\includegraphics[width=0.78\linewidth,angle=0]{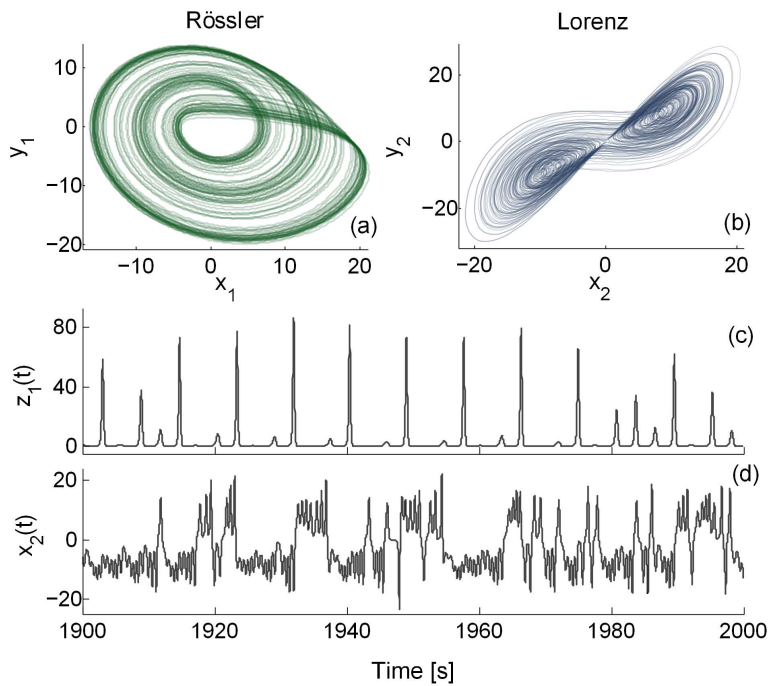}
\caption{\label{fig6:chaos} The coupled chaotic R\"{o}ssler and Lorenz systems with time-varying parameters and subject to noise (\ref{eq:chaos1}),(\ref{eq:chaos2}). The phase portraits or strange attractors are shown (a) for the R\"{o}ssler and (b) for the Lorenz systems. Short time-segments of the chaotic time series are presented: (c) $z_1(t)$ from the R\"{o}ssler and (d) $x_2(t)$ from the Lorenz system.}
\end{center}
\end{figure}

It was at the beginning of the chaotic communication era when P\'{e}rez and Cerdeira \cite{Perez:95} demonstrated a way of breaking such communication schemes by extracting the messages they carried. Their methods were based on reconstruction of the attractors' properties from the transmitted signals only. This valuable work not only broke the existing schemes, but also introduced an attacking principle that all future (chaotic) attractor-based schemes must resist. The new scheme based on coupling functions and dynamical Bayesian inference is able to withstand such attacks because the information is encrypted in the weak coupling between two independent and self-sustained attractors.

To illustrate dynamical Bayesian inference in state space we consider a coupled pair of chaotic R\"{o}ssler and Lorenz oscillators. This coupled system represents the coupling-function-based communication model and is duplicated on both the transmitter and receiver sides. The model is given by a R\"{o}ssler system:
\begin{equation}
\begin{split}
\dot x_1=& -2y_1-z_1\\
\dot y_1=& 2x_1+0.45y_1+\xi_1(t)\\
\dot z_1=& 2+x_1z_1-10z_1,\text{     }\text{     }\text{     }\text{     }\text{     }\text{     }\text{     }\text{     }\text{     }\text{     } \text{     } \text{     } \text{     }\text{     }\text{     }\text{     }\text{     } \text{     } \text{     } \text{     }\text{     }\text{     } \text{     }\text{     }\text{     }\text{     }\text{     } \text{     } \text{     } \text{     }\text{     } \text{     } \text{     }\text{     }\text{     }\text{     }\text{     } \text{     } \text{     } \text{     }\text{     } \text{     } \text{     }\text{     }\text{     }\text{     }\text{     }   \\
\end{split}
\label{eq:chaos1}
\end{equation}
driving a Lorenz system
\begin{equation}
\begin{split}
 \dot x_2=& 10y_2-10x_2\\
 \dot y_2=& 28 x_2-x_2z_2-y_2\\
 \dot z_2=& x_2y_2-2.66z_2+\varepsilon_1(t)y_1+\varepsilon_2(t)x_1z_1+\xi_2(t),
\end{split}
\label{eq:chaos2}
\end{equation}
where the noises are taken to be white and Gaussian with noise intensities $E_1=0.05$ and $E_2=0.3$. The coupling parameters $\varepsilon_1(t)$ and $\varepsilon_2(t)$ are set to be time-varying i.e.\ they can represent the information messages that are to be securely encrypted. The first coupling is binary $\varepsilon_1(t)=\{0,2\}$, while the second is continuous $\varepsilon_2(t)=3+0.3\sin(2\pi 0.001t)$. The attractors and signals of such chaotic systems subject to noise and time-varying couplings are shown in Fig.\ \ref{fig6:chaos}.

\begin{figure}[t]
\begin{center}
\includegraphics[width=0.82\linewidth,angle=0]{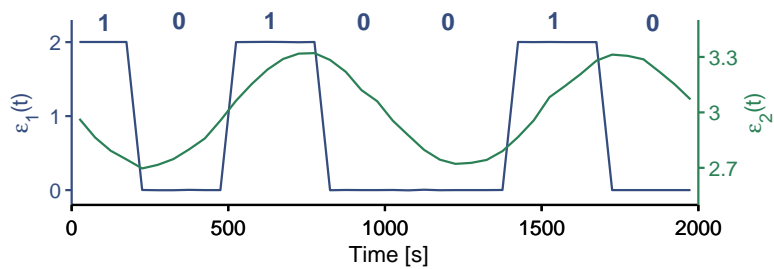}
\caption{\label{fig7:comm} The inferred time-varying coupling parameters of the model (\ref{eq:chaos1}),(\ref{eq:chaos2}) as an illustration of the information carriers in secure communications. The ordinate axis on the left shows the binary time-varying parameter $\varepsilon_1(t)$, which can be used e.g.\ to encrypt the digital message \emph{1010010} shown along the top of the figure. One can also encrypt another signal simultaneously, as illustrated by the continuous sine message embodied as the variation of the $\varepsilon_2(t)$ parameter on the right ordinate axis.}
\end{center}
\end{figure}

This particular example is very convenient for the use of dynamical Bayesian inference on the receiver side because one knows both the model and its base functions \emph{a priori}, and has access to the time-series of all the dimensions. Additionally, the chaotic signals with their notion of strange-attractors tend to span a broad region of state space (larger than e.g.\ limit-cycle oscillators), which provides more information and makes the inference easier and more precise. Hence, by using the functions on the \emph{rhs} of systems (\ref{eq:chaos1}),(\ref{eq:chaos2}) as base functions, one can infer the model parameters and their time-variability. Among them are the two time-varying coupling parameters, which in the communication example could convey the information messages as shown in Fig.\ \ref{fig7:comm}. As it can be seen, both the digital (e.g.\ $1010010$) and the continuous (e.g.\ sine or speech) message can be inferred with great precision.

\section{Discussion and conclusion}
\label{s:conclusion}

The aim of this review has been to provide insights into the implementation of dynamical Bayesian inference in a clear and simple way. Further details of the method and its applications can be find in \cite{Stankovski:12b,Duggento:12,Luchinsky:08,Duggento:08,Smelyanskiy:05a}. Needless to say, the overall framework can be much broader and a number of important generalizations are possible.

The examples here included only two coupled systems, while in general the technique can be applied to a larger group e.g.\ a small-scale network of oscillators \cite{Duggento:12,Kralemann:11}. The phase decompositions can be applied for pairwise couplings but, more importantly, joint coupling influences can also be inferred. In such cases one can detect unique states characteristic only of the joint couplings, like triplet synchronization \cite{Kralemann:13}. In these kind of analysis, the effective coupling \cite{Friston:11,Park:13} should be distinguished by appropriate use of surrogate testing \cite{Schreiber:00b}.

The procedure of information propagation that allows time-variability to be tracked, depends on $p_w$ which acts as a free parameter. One can further improve this procedure by making the parameter adaptive in order to follow the time-variability more closely and to infer the noise more precisely. This might be realized by determining the optimal parameter from a spectrum of values within each window. Even though this procedure would be very slow, it might prove helpful and could be necessary in certain cases.

The phase base functions are not strictly confined all to be from Fourier series, and other additional functions can be included. For example, if we have expansion up to second order $K=2$, one can include also other components such as $\sin(7\phi_1-4\phi_2)$ in order to detect synchronization more precisely for 7:4 synchronization ratios. Similarly the state base functions can be extended to include a large set of functions and only the ones that intrinsically belong to the underlying model will be inferred as `non-zero' \cite{Smelyanskiy:05a}.

Within the framework of dynamical inference, the differential $\dot \cci_{n}$ is calculated as $\dot \cci_{n}=({\cci}_{{n+1}}-\cci_{n})/{h}$. Improved performance can be accomplished if one provides this as directly estimated instantaneous frequency -- e.g.\ using the synchrosqueezed transform \cite{Daubechies:11} or nonlinear mode decomposition \cite{Iatsenko:13a}. Alternatively, certain digital filters (e.g.\ the Savitzky-Golay filter) can be used for smoothing the noise effect before evaluating the derivatives.

The inference of causality between interacting oscillatory systems has attracted much attention in the last decade. In addition to nonlinear dynamics methods, both for inferring direction of coupling or the coupling function \cite{Rosenblum:01,Kralemann:13b}, methods based on information theory \cite{Palus:03a} and wavelet bispectral analysis \cite{Jamsek:10}, there is the dynamical Bayesian inference method presented above. It facilitates comprehensive reconstruction of the dynamical properties of the interacting systems -- either two or a whole network -- and allows every aspect of their interactions to be studied, including their synchronization, direction of coupling, and coupling function. The approach provides deep insight into the properties of the dynamical systems of interest and thus makes possible both diagnosis and prognosis of their behaviour.

Of course, dynamical Bayesian inference possesses relationships, similarities and complementarities with other methods, including the inference of deterministic or stochastic models, based on Bayesian theory, particle filters and maximum likelihood estimators \cite{Smirnov:09,Friston:02,Arulampalam:02,Voss:04,Kiss:05}. The inference of coupling causality has huge practical applicability and methods based on Granger causality or transfer entropy  \cite{Granger:69,Seth:05,Porta:14,Schreiber:00a,Palus:03a} have recently become popular in this context. We note however, that such methods infer statistical effects, while the method presented here, being based on a dynamical model, can infer causal mechanisms \cite{Barrett:13}. In other words, Granger causality-like methods infer only the existence of causal effect and not the nature and mechanism of the cause itself. In neuroscience, Granger causality methods are linked to directed functional connectivity, while dynamical inference methods distinguish effective connectivity \cite{Friston:11}.

In summary, this tutorial has been intended to familiarize the reader with a technique for the Bayesian inference of time-evolving coupled dynamics in the presence of noise. A comprehensive description of the method, including the theoretical constraints, algorithms, implementation, and demonstrations of their main features in relation to a few characteristic examples has been provided. To facilitate the first steps in applying this powerful and useful method MatLab codes including examples are also being made available. We hope that the tutorial will lead the reader to new insights into dynamical phenomena from measured data, and that it will provide a useful aid for tackling a diverse range of signal processing problems.

\section{Acknowledgments}
Our grateful thanks are due to Dmytro Iatsenko, Valentina Ticcinelli, Will Gibby, Gemma Lancaster, Phil Clemson and Yevhen Suprunenko for testing the software codes and for the  useful comments on the manuscript. This work was supported by the Engineering and Physical Sciences Research Council (UK) [Grant No. EP/100999X1].

\bibliographystyle{epj}

\end{document}